\documentclass[runningheads]{llncs}
\usepackage{graphicx}

\usepackage{booktabs,makecell} 
\usepackage{booktabs}
\usepackage[utf8]{inputenc}
\usepackage{amsmath}
\usepackage{graphicx}
\usepackage{wrapfig}
\usepackage{float}
\usepackage{arydshln}
\usepackage{xspace}
\usepackage{algorithmic}
\usepackage[ruled]{algorithm2e}
\makeatletter
\newcommand{\removelatexerror}{\let\@latex@error\@gobble}
\makeatother
\usepackage{amsmath,amssymb,amsfonts}
\usepackage{graphicx}
\usepackage[labelformat = parens,justification=centering]{subfig}
\usepackage{textcomp}
\usepackage{xcolor}
\usepackage[normalem]{ulem}
\usepackage{todonotes}
\usepackage{multirow}
\usepackage[gen]{eurosym}
\usepackage{placeins}
\usepackage{tabularx}

\definecolor{amblu}{RGB}{55, 126, 184}
\definecolor{gray30}{rgb}{0.3,0.3,0.3}
\usepackage{listings}
\usepackage{xspace}

\newcommand{\lcudasyntex}{\lstinline[columns=fixed]{<<<}\xspace}
\newcommand{\rcudasyntex}{\lstinline[columns=fixed]{>>>}\xspace}
\newcommand{\launchbound}{\texttt{\_\_launch\_bounds\_\_}\xspace}
\newcommand{\kernel}{\texttt{\_\_global\_\_}\xspace}

\newcommand{\gko}{\textsc{Ginkgo}\xspace}

\newcommand{\lstbg}[3][0pt]{{\fboxsep#1\colorbox{#2}{\strut #3}}}
\lstdefinelanguage{diff}{
  basicstyle=\scriptsize\tt,
  numbers=left,                   
  numberstyle=\scriptsize,      
  stepnumber=1,                   
  numbersep=5pt,
  stringstyle=\scriptsize\tt,
  identifierstyle=\scriptsize\tt,
  commentstyle=\scriptsize\em,
  keywordstyle=\scriptsize\bf,
  breaklines,frameround=ffff,frame=trbl,captionpos=b,
  morecomment=[f][\lstbg{red!20}]-,
  morecomment=[f][\lstbg{green!20}]+,
  morecomment=[f][\textit]{@@}
}

\lstdefinelanguage{CMake}
	{
    basicstyle=\scriptsize\tt,
    numbers=left,                   
    numberstyle=\scriptsize,      
    stepnumber=1,                   
    numbersep=5pt,
    stringstyle=\scriptsize\tt,
    identifierstyle=\scriptsize\tt,
    commentstyle=\scriptsize\em,
    keywordstyle=\scriptsize\bf,
    breaklines,frameround=ffff,frame=trbl,captionpos=b,
    morekeywords={
add_custom_command,
add_custom_target,
add_definitions,
add_dependencies,
add_executable,
add_library,
add_subdirectory,
add_test,
aux_source_directory,
break,
build_command,
cmake_minimum_required,
cmake_policy,
configure_file,
create_test_sourcelist,
define_property,
else,
elseif,
enable_language,
enable_testing,
endforeach,
endfunction,
endif,
endmacro,
endwhile,
execute_process,
export,
file,
find_file,
find_library,
find_package,
find_path,
find_program,
fltk_wrap_ui,
foreach,
function,
get_cmake_property,
get_directory_property,
get_filename_component,
get_property,
get_source_file_property,
get_target_property,
get_test_property,
if,
include,
include_directories,
include_external_msproject,
include_regular_expression,
install,
link_directories,
list,
load_cache,
load_command,
macro,
mark_as_advanced,
math,
message,
option,
project,
qt_wrap_cpp,
qt_wrap_ui,
remove_definitions,
return,
separate_arguments,
set,
set_directory_properties,
set_property,
set_source_files_properties,
set_target_properties,
set_tests_properties,
site_name,
source_group,
string,
target_link_libraries,
try_compile,
try_run,
unset,
variable_watch,
while,
build_name,
exec_program,
export_library_dependencies,
install_files,
install_programs,
install_targets,
link_libraries,
make_directory,
output_required_files,
remove,
subdir_depends,
subdirs,
use_mangled_mesa,
utility_source,
variable_requires,
write_file,
READ, WRITE, APPEND, RENAME, DOWNLOAD, UPLOAD,
GLOB, GLOB_RECURSE, MAKE_DIRECTORY,
TO_CMAKE_PATH, TO_NATIVE_PATH,
LENGTH,GET,FIND, APPEND, INSERT, REMOVE_ITEM, REMOVE_AT, REMOVE_DUPLICATES, REVERSE, SORT,
STATUS, WARNING, LOG, SHOW_PROGRESS, EXISTS, COMMAND,
RESULT_VARIABLE, OUTPUT_VARIABLE, ERROR_VARIABLE,
PROPERTIES
         },
         morekeywords=[2]{
<PROJECT-NAME>_BINARY_DIR,
<PROJECT-NAME>_SOURCE_DIR,
<PROJECT-NAME>_VERSION,
<PROJECT-NAME>_VERSION_MAJOR,
<PROJECT-NAME>_VERSION_MINOR,
<PROJECT-NAME>_VERSION_PATCH,
<PROJECT-NAME>_VERSION_TWEAK,
APPLE,
BORLAND,
BUILD_SHARED_LIBS,
CMAKE_<CONFIG>_POSTFIX,
CMAKE_<LANG>_ARCHIVE_APPEND,
CMAKE_<LANG>_ARCHIVE_CREATE,
CMAKE_<LANG>_ARCHIVE_FINISH,
CMAKE_<LANG>_CLANG_TIDY,
CMAKE_<LANG>_COMPILER,
CMAKE_<LANG>_COMPILER_ABI,
CMAKE_<LANG>_COMPILER_EXTERNAL_TOOLCHAIN,
CMAKE_<LANG>_COMPILER_ID,
CMAKE_<LANG>_COMPILER_LAUNCHER,
CMAKE_<LANG>_COMPILER_LOADED,
CMAKE_<LANG>_COMPILER_TARGET,
CMAKE_<LANG>_COMPILER_VERSION,
CMAKE_<LANG>_COMPILE_OBJECT,
CMAKE_<LANG>_CREATE_SHARED_LIBRARY,
CMAKE_<LANG>_CREATE_SHARED_MODULE,
CMAKE_<LANG>_CREATE_STATIC_LIBRARY,
CMAKE_<LANG>_FLAGS,
CMAKE_<LANG>_FLAGS_DEBUG,
CMAKE_<LANG>_FLAGS_MINSIZEREL,
CMAKE_<LANG>_FLAGS_RELEASE,
CMAKE_<LANG>_FLAGS_RELWITHDEBINFO,
CMAKE_<LANG>_GHS_KERNEL_FLAGS_DEBUG,
CMAKE_<LANG>_GHS_KERNEL_FLAGS_MINSIZEREL,
CMAKE_<LANG>_GHS_KERNEL_FLAGS_RELEASE,
CMAKE_<LANG>_GHS_KERNEL_FLAGS_RELWITHDEBINFO,
CMAKE_<LANG>_IGNORE_EXTENSIONS,
CMAKE_<LANG>_IMPLICIT_INCLUDE_DIRECTORIES,
CMAKE_<LANG>_IMPLICIT_LINK_DIRECTORIES,
CMAKE_<LANG>_IMPLICIT_LINK_FRAMEWORK_DIRECTORIES,
CMAKE_<LANG>_IMPLICIT_LINK_LIBRARIES,
CMAKE_<LANG>_INCLUDE_WHAT_YOU_USE,
CMAKE_<LANG>_LIBRARY_ARCHITECTURE,
CMAKE_<LANG>_LINKER_PREFERENCE,
CMAKE_<LANG>_LINKER_PREFERENCE_PROPAGATES,
CMAKE_<LANG>_LINK_EXECUTABLE,
CMAKE_<LANG>_OUTPUT_EXTENSION,
CMAKE_<LANG>_PLATFORM_ID,
CMAKE_<LANG>_SIMULATE_ID,
CMAKE_<LANG>_SIMULATE_VERSION,
CMAKE_<LANG>_SIZEOF_DATA_PTR,
CMAKE_<LANG>_SOURCE_FILE_EXTENSIONS,
CMAKE_<LANG>_STANDARD_INCLUDE_DIRECTORIES,
CMAKE_<LANG>_STANDARD_LIBRARIES,
CMAKE_<LANG>_VISIBILITY_PRESET,
CMAKE_ABSOLUTE_DESTINATION_FILES,
CMAKE_ANDROID_ANT_ADDITIONAL_OPTIONS,
CMAKE_ANDROID_API,
CMAKE_ANDROID_API_MIN,
CMAKE_ANDROID_ARCH,
CMAKE_ANDROID_ASSETS_DIRECTORIES,
CMAKE_ANDROID_GUI,
CMAKE_ANDROID_JAR_DEPENDENCIES,
CMAKE_ANDROID_JAR_DIRECTORIES,
CMAKE_ANDROID_JAVA_SOURCE_DIR,
CMAKE_ANDROID_NATIVE_LIB_DEPENDENCIES,
CMAKE_ANDROID_NATIVE_LIB_DIRECTORIES,
CMAKE_ANDROID_PROCESS_MAX,
CMAKE_ANDROID_PROGUARD,
CMAKE_ANDROID_PROGUARD_CONFIG_PATH,
CMAKE_ANDROID_SECURE_PROPS_PATH,
CMAKE_ANDROID_SKIP_ANT_STEP,
CMAKE_ANDROID_STL_TYPE,
CMAKE_APPBUNDLE_PATH,
CMAKE_AR,
CMAKE_ARCHIVE_OUTPUT_DIRECTORY,
CMAKE_ARCHIVE_OUTPUT_DIRECTORY_<CONFIG>,
CMAKE_ARGC,
CMAKE_ARGV0,
CMAKE_AUTOMOC,
CMAKE_AUTOMOC_MOC_OPTIONS,
CMAKE_AUTOMOC_RELAXED_MODE,
CMAKE_AUTORCC,
CMAKE_AUTORCC_OPTIONS,
CMAKE_AUTOUIC,
CMAKE_AUTOUIC_OPTIONS,
CMAKE_BACKWARDS_COMPATIBILITY,
CMAKE_BINARY_DIR,
CMAKE_BUILD_TOOL,
CMAKE_BUILD_TYPE,
CMAKE_BUILD_WITH_INSTALL_RPATH,
CMAKE_CACHEFILE_DIR,
CMAKE_CACHE_MAJOR_VERSION,
CMAKE_CACHE_MINOR_VERSION,
CMAKE_CACHE_PATCH_VERSION,
CMAKE_CFG_INTDIR,
CMAKE_CL_64,
CMAKE_COLOR_MAKEFILE,
CMAKE_COMMAND,
CMAKE_COMPILER_2005,
CMAKE_COMPILER_IS_GNU<LANG>,
CMAKE_COMPILE_PDB_OUTPUT_DIRECTORY,
CMAKE_COMPILE_PDB_OUTPUT_DIRECTORY_<CONFIG>,
CMAKE_CONFIGURATION_TYPES,
CMAKE_CROSSCOMPILING,
CMAKE_CROSSCOMPILING_EMULATOR,
CMAKE_CTEST_COMMAND,
CMAKE_CURRENT_BINARY_DIR,
CMAKE_CURRENT_LIST_DIR,
CMAKE_CURRENT_LIST_FILE,
CMAKE_CURRENT_LIST_LINE,
CMAKE_CURRENT_SOURCE_DIR,
CMAKE_CXX_COMPILE_FEATURES,
CMAKE_CXX_EXTENSIONS,
CMAKE_CXX_STANDARD,
CMAKE_CXX_STANDARD_REQUIRED,
CMAKE_C_COMPILE_FEATURES,
CMAKE_C_EXTENSIONS,
CMAKE_C_STANDARD,
CMAKE_C_STANDARD_REQUIRED,
CMAKE_DEBUG_POSTFIX,
CMAKE_DEBUG_TARGET_PROPERTIES,
CMAKE_DEPENDS_IN_PROJECT_ONLY,
CMAKE_DISABLE_FIND_PACKAGE_<PackageName>,
CMAKE_DL_LIBS,
CMAKE_ECLIPSE_GENERATE_LINKED_RESOURCES,
CMAKE_ECLIPSE_GENERATE_SOURCE_PROJECT,
CMAKE_ECLIPSE_MAKE_ARGUMENTS,
CMAKE_ECLIPSE_VERSION,
CMAKE_EDIT_COMMAND,
CMAKE_ENABLE_EXPORTS,
CMAKE_ERROR_DEPRECATED,
CMAKE_ERROR_ON_ABSOLUTE_INSTALL_DESTINATION,
CMAKE_EXECUTABLE_SUFFIX,
CMAKE_EXE_LINKER_FLAGS,
CMAKE_EXE_LINKER_FLAGS_<CONFIG>,
CMAKE_EXPORT_COMPILE_COMMANDS,
CMAKE_EXPORT_NO_PACKAGE_REGISTRY,
CMAKE_EXTRA_GENERATOR,
CMAKE_EXTRA_SHARED_LIBRARY_SUFFIXES,
CMAKE_FIND_APPBUNDLE,
CMAKE_FIND_FRAMEWORK,
CMAKE_FIND_LIBRARY_PREFIXES,
CMAKE_FIND_LIBRARY_SUFFIXES,
CMAKE_FIND_NO_INSTALL_PREFIX,
CMAKE_FIND_PACKAGE_NAME,
CMAKE_FIND_PACKAGE_NO_PACKAGE_REGISTRY,
CMAKE_FIND_PACKAGE_NO_SYSTEM_PACKAGE_REGISTRY,
CMAKE_FIND_PACKAGE_WARN_NO_MODULE,
CMAKE_FIND_ROOT_PATH,
CMAKE_FIND_ROOT_PATH_MODE_INCLUDE,
CMAKE_FIND_ROOT_PATH_MODE_LIBRARY,
CMAKE_FIND_ROOT_PATH_MODE_PACKAGE,
CMAKE_FIND_ROOT_PATH_MODE_PROGRAM,
CMAKE_FRAMEWORK_PATH,
CMAKE_Fortran_FORMAT,
CMAKE_Fortran_MODDIR_DEFAULT,
CMAKE_Fortran_MODDIR_FLAG,
CMAKE_Fortran_MODOUT_FLAG,
CMAKE_Fortran_MODULE_DIRECTORY,
CMAKE_GENERATOR,
CMAKE_GENERATOR_PLATFORM,
CMAKE_GENERATOR_TOOLSET,
CMAKE_GNUtoMS,
CMAKE_HOME_DIRECTORY,
CMAKE_HOST_APPLE,
CMAKE_HOST_SOLARIS,
CMAKE_HOST_SYSTEM,
CMAKE_HOST_SYSTEM_NAME,
CMAKE_HOST_SYSTEM_PROCESSOR,
CMAKE_HOST_SYSTEM_VERSION,
CMAKE_HOST_UNIX,
CMAKE_HOST_WIN32,
CMAKE_IGNORE_PATH,
CMAKE_IMPORT_LIBRARY_PREFIX,
CMAKE_IMPORT_LIBRARY_SUFFIX,
CMAKE_INCLUDE_CURRENT_DIR,
CMAKE_INCLUDE_CURRENT_DIR_IN_INTERFACE,
CMAKE_INCLUDE_DIRECTORIES_BEFORE,
CMAKE_INCLUDE_DIRECTORIES_PROJECT_BEFORE,
CMAKE_INCLUDE_PATH,
CMAKE_INSTALL_DEFAULT_COMPONENT_NAME,
CMAKE_INSTALL_MESSAGE,
CMAKE_INSTALL_NAME_DIR,
CMAKE_INSTALL_PREFIX,
CMAKE_INSTALL_RPATH,
CMAKE_INSTALL_RPATH_USE_LINK_PATH,
CMAKE_INTERNAL_PLATFORM_ABI,
CMAKE_IOS_INSTALL_COMBINED,
CMAKE_JOB_POOL_COMPILE,
CMAKE_JOB_POOL_LINK,
CMAKE_LIBRARY_ARCHITECTURE,
CMAKE_LIBRARY_ARCHITECTURE_REGEX,
CMAKE_LIBRARY_OUTPUT_DIRECTORY,
CMAKE_LIBRARY_OUTPUT_DIRECTORY_<CONFIG>,
CMAKE_LIBRARY_PATH,
CMAKE_LIBRARY_PATH_FLAG,
CMAKE_LINK_DEF_FILE_FLAG,
CMAKE_LINK_DEPENDS_NO_SHARED,
CMAKE_LINK_INTERFACE_LIBRARIES,
CMAKE_LINK_LIBRARY_FILE_FLAG,
CMAKE_LINK_LIBRARY_FLAG,
CMAKE_LINK_LIBRARY_SUFFIX,
CMAKE_LINK_SEARCH_END_STATIC,
CMAKE_LINK_SEARCH_START_STATIC,
CMAKE_MACOSX_BUNDLE,
CMAKE_MACOSX_RPATH,
CMAKE_MAJOR_VERSION,
CMAKE_MAKE_PROGRAM,
CMAKE_MAP_IMPORTED_CONFIG_<CONFIG>,
CMAKE_MATCH_COUNT,
CMAKE_MFC_FLAG,
CMAKE_MINIMUM_REQUIRED_VERSION,
CMAKE_MINOR_VERSION,
CMAKE_MODULE_LINKER_FLAGS,
CMAKE_MODULE_LINKER_FLAGS_<CONFIG>,
CMAKE_MODULE_PATH,
CMAKE_NINJA_OUTPUT_PATH_PREFIX,
CMAKE_NOT_USING_CONFIG_FLAGS,
CMAKE_NO_BUILTIN_CHRPATH,
CMAKE_NO_SYSTEM_FROM_IMPORTED,
CMAKE_OBJECT_PATH_MAX,
CMAKE_OSX_ARCHITECTURES,
CMAKE_OSX_DEPLOYMENT_TARGET,
CMAKE_OSX_SYSROOT,
CMAKE_PARENT_LIST_FILE,
CMAKE_PATCH_VERSION,
CMAKE_PDB_OUTPUT_DIRECTORY,
CMAKE_PDB_OUTPUT_DIRECTORY_<CONFIG>,
CMAKE_POLICY_DEFAULT_CMP<NNNN>,
CMAKE_POLICY_WARNING_CMP<NNNN>,
CMAKE_POSITION_INDEPENDENT_CODE,
CMAKE_PREFIX_PATH,
CMAKE_PROGRAM_PATH,
CMAKE_PROJECT_<PROJECT-NAME>_INCLUDE,
CMAKE_PROJECT_NAME,
CMAKE_RANLIB,
CMAKE_ROOT,
CMAKE_RUNTIME_OUTPUT_DIRECTORY,
CMAKE_RUNTIME_OUTPUT_DIRECTORY_<CONFIG>,
CMAKE_SCRIPT_MODE_FILE,
CMAKE_SHARED_LIBRARY_PREFIX,
CMAKE_SHARED_LIBRARY_SUFFIX,
CMAKE_SHARED_LINKER_FLAGS,
CMAKE_SHARED_LINKER_FLAGS_<CONFIG>,
CMAKE_SHARED_MODULE_PREFIX,
CMAKE_SHARED_MODULE_SUFFIX,
CMAKE_SIZEOF_VOID_P,
CMAKE_SKIP_BUILD_RPATH,
CMAKE_SKIP_INSTALL_ALL_DEPENDENCY,
CMAKE_SKIP_INSTALL_RPATH,
CMAKE_SKIP_INSTALL_RULES,
CMAKE_SKIP_RPATH,
CMAKE_SOURCE_DIR,
CMAKE_STAGING_PREFIX,
CMAKE_STATIC_LIBRARY_PREFIX,
CMAKE_STATIC_LIBRARY_SUFFIX,
CMAKE_STATIC_LINKER_FLAGS,
CMAKE_STATIC_LINKER_FLAGS_<CONFIG>,
CMAKE_SYSROOT,
CMAKE_SYSTEM,
CMAKE_SYSTEM_APPBUNDLE_PATH,
CMAKE_SYSTEM_FRAMEWORK_PATH,
CMAKE_SYSTEM_IGNORE_PATH,
CMAKE_SYSTEM_INCLUDE_PATH,
CMAKE_SYSTEM_LIBRARY_PATH,
CMAKE_SYSTEM_NAME,
CMAKE_SYSTEM_PREFIX_PATH,
CMAKE_SYSTEM_PROCESSOR,
CMAKE_SYSTEM_PROGRAM_PATH,
CMAKE_SYSTEM_VERSION,
CMAKE_TOOLCHAIN_FILE,
CMAKE_TRY_COMPILE_CONFIGURATION,
CMAKE_TRY_COMPILE_PLATFORM_VARIABLES,
CMAKE_TRY_COMPILE_TARGET_TYPE,
CMAKE_TWEAK_VERSION,
CMAKE_USER_MAKE_RULES_OVERRIDE,
CMAKE_USER_MAKE_RULES_OVERRIDE_<LANG>,
CMAKE_USE_RELATIVE_PATHS,
CMAKE_VERBOSE_MAKEFILE,
CMAKE_VERSION,
CMAKE_VISIBILITY_INLINES_HIDDEN,
CMAKE_VS_DEVENV_COMMAND,
CMAKE_VS_INCLUDE_INSTALL_TO_DEFAULT_BUILD,
CMAKE_VS_INTEL_Fortran_PROJECT_VERSION,
CMAKE_VS_MSBUILD_COMMAND,
CMAKE_VS_NsightTegra_VERSION,
CMAKE_VS_PLATFORM_NAME,
CMAKE_VS_PLATFORM_TOOLSET,
CMAKE_VS_WINDOWS_TARGET_PLATFORM_VERSION,
CMAKE_WARN_DEPRECATED,
CMAKE_WARN_ON_ABSOLUTE_INSTALL_DESTINATION,
CMAKE_WIN32_EXECUTABLE,
CMAKE_WINDOWS_EXPORT_ALL_SYMBOLS,
CMAKE_XCODE_ATTRIBUTE_<an-attribute>,
CMAKE_XCODE_PLATFORM_TOOLSET,
CPACK_ABSOLUTE_DESTINATION_FILES,
CPACK_COMPONENT_INCLUDE_TOPLEVEL_DIRECTORY,
CPACK_ERROR_ON_ABSOLUTE_INSTALL_DESTINATION,
CPACK_INCLUDE_TOPLEVEL_DIRECTORY,
CPACK_INSTALL_SCRIPT,
CPACK_PACKAGING_INSTALL_PREFIX,
CPACK_SET_DESTDIR,
CPACK_WARN_ON_ABSOLUTE_INSTALL_DESTINATION,
CTEST_BINARY_DIRECTORY,
CTEST_BUILD_COMMAND,
CTEST_BUILD_NAME,
CTEST_BZR_COMMAND,
CTEST_BZR_UPDATE_OPTIONS,
CTEST_CHANGE_ID,
CTEST_CHECKOUT_COMMAND,
CTEST_CONFIGURATION_TYPE,
CTEST_CONFIGURE_COMMAND,
CTEST_COVERAGE_COMMAND,
CTEST_COVERAGE_EXTRA_FLAGS,
CTEST_CURL_OPTIONS,
CTEST_CUSTOM_COVERAGE_EXCLUDE,
CTEST_CUSTOM_ERROR_EXCEPTION,
CTEST_CUSTOM_ERROR_MATCH,
CTEST_CUSTOM_ERROR_POST_CONTEXT,
CTEST_CUSTOM_ERROR_PRE_CONTEXT,
CTEST_CUSTOM_MAXIMUM_FAILED_TEST_OUTPUT_SIZE,
CTEST_CUSTOM_MAXIMUM_NUMBER_OF_ERRORS,
CTEST_CUSTOM_MAXIMUM_NUMBER_OF_WARNINGS,
CTEST_CUSTOM_MAXIMUM_PASSED_TEST_OUTPUT_SIZE,
CTEST_CUSTOM_MEMCHECK_IGNORE,
CTEST_CUSTOM_POST_MEMCHECK,
CTEST_CUSTOM_POST_TEST,
CTEST_CUSTOM_PRE_MEMCHECK,
CTEST_CUSTOM_PRE_TEST,
CTEST_CUSTOM_TEST_IGNORE,
CTEST_CUSTOM_WARNING_EXCEPTION,
CTEST_CUSTOM_WARNING_MATCH,
CTEST_CVS_CHECKOUT,
CTEST_CVS_COMMAND,
CTEST_CVS_UPDATE_OPTIONS,
CTEST_DROP_LOCATION,
CTEST_DROP_METHOD,
CTEST_DROP_SITE,
CTEST_DROP_SITE_CDASH,
CTEST_DROP_SITE_PASSWORD,
CTEST_DROP_SITE_USER,
CTEST_EXTRA_COVERAGE_GLOB,
CTEST_GIT_COMMAND,
CTEST_GIT_INIT_SUBMODULES,
CTEST_GIT_UPDATE_CUSTOM,
CTEST_GIT_UPDATE_OPTIONS,
CTEST_HG_COMMAND,
CTEST_HG_UPDATE_OPTIONS,
CTEST_MEMORYCHECK_COMMAND,
CTEST_MEMORYCHECK_COMMAND_OPTIONS,
CTEST_MEMORYCHECK_SANITIZER_OPTIONS,
CTEST_MEMORYCHECK_SUPPRESSIONS_FILE,
CTEST_MEMORYCHECK_TYPE,
CTEST_NIGHTLY_START_TIME,
CTEST_P4_CLIENT,
CTEST_P4_COMMAND,
CTEST_P4_OPTIONS,
CTEST_P4_UPDATE_OPTIONS,
CTEST_SCP_COMMAND,
CTEST_SITE,
CTEST_SOURCE_DIRECTORY,
CTEST_SVN_COMMAND,
CTEST_SVN_OPTIONS,
CTEST_SVN_UPDATE_OPTIONS,
CTEST_TEST_LOAD,
CTEST_TEST_TIMEOUT,
CTEST_TRIGGER_SITE,
CTEST_UPDATE_COMMAND,
CTEST_UPDATE_OPTIONS,
CTEST_UPDATE_VERSION_ONLY,
CTEST_USE_LAUNCHERS,
CYGWIN,
ENV,
EXECUTABLE_OUTPUT_PATH,
GHS-MULTI,
LIBRARY_OUTPUT_PATH,
MINGW,
MSVC,
MSVC10,
MSVC11,
MSVC12,
MSVC14,
MSVC60,
MSVC70,
MSVC71,
MSVC80,
MSVC90,
MSVC_IDE,
MSVC_VERSION,
PROJECT_BINARY_DIR,
PROJECT_NAME,
PROJECT_SOURCE_DIR,
PROJECT_VERSION,
PROJECT_VERSION_MAJOR,
PROJECT_VERSION_MINOR,
PROJECT_VERSION_PATCH,
PROJECT_VERSION_TWEAK,
UNIX,
WIN32,
WINCE,
WINDOWS_PHONE,
WINDOWS_STORE,
XCODE_VERSION
         },
         morekeywords=[3]{
<CONFIG>_OUTPUT_NAME,
<CONFIG>_POSTFIX,
<LANG>_CLANG_TIDY,
<LANG>_COMPILER_LAUNCHER,
<LANG>_INCLUDE_WHAT_YOU_USE,
<LANG>_VISIBILITY_PRESET,
ABSTRACT,
ADDITIONAL_MAKE_CLEAN_FILES,
ADVANCED,
ALIASED_TARGET,
ALLOW_DUPLICATE_CUSTOM_TARGETS,
ANDROID_ANT_ADDITIONAL_OPTIONS,
ANDROID_API,
ANDROID_API_MIN,
ANDROID_ARCH,
ANDROID_ASSETS_DIRECTORIES,
ANDROID_GUI,
ANDROID_JAR_DEPENDENCIES,
ANDROID_JAR_DIRECTORIES,
ANDROID_JAVA_SOURCE_DIR,
ANDROID_NATIVE_LIB_DEPENDENCIES,
ANDROID_NATIVE_LIB_DIRECTORIES,
ANDROID_PROCESS_MAX,
ANDROID_PROGUARD,
ANDROID_PROGUARD_CONFIG_PATH,
ANDROID_SECURE_PROPS_PATH,
ANDROID_SKIP_ANT_STEP,
ANDROID_STL_TYPE,
ARCHIVE_OUTPUT_DIRECTORY,
ARCHIVE_OUTPUT_DIRECTORY_<CONFIG>,
ARCHIVE_OUTPUT_NAME,
ARCHIVE_OUTPUT_NAME_<CONFIG>,
ATTACHED_FILES,
ATTACHED_FILES_ON_FAIL,
AUTOGEN_TARGETS_FOLDER,
AUTOGEN_TARGET_DEPENDS,
AUTOMOC,
AUTOMOC_MOC_OPTIONS,
AUTOMOC_TARGETS_FOLDER,
AUTORCC,
AUTORCC_OPTIONS,
AUTORCC_OPTIONS,
AUTOUIC,
AUTOUIC_OPTIONS,
AUTOUIC_OPTIONS,
BINARY_DIR,
BUILD_WITH_INSTALL_RPATH,
BUNDLE,
BUNDLE_EXTENSION,
CACHE_VARIABLES,
CLEAN_NO_CUSTOM,
CMAKE_CONFIGURE_DEPENDS,
CMAKE_CXX_KNOWN_FEATURES,
CMAKE_C_KNOWN_FEATURES,
COMPATIBLE_INTERFACE_BOOL,
COMPATIBLE_INTERFACE_NUMBER_MAX,
COMPATIBLE_INTERFACE_NUMBER_MIN,
COMPATIBLE_INTERFACE_STRING,
COMPILE_DEFINITIONS,
COMPILE_DEFINITIONS,
COMPILE_DEFINITIONS,
COMPILE_DEFINITIONS_<CONFIG>,
COMPILE_DEFINITIONS_<CONFIG>,
COMPILE_DEFINITIONS_<CONFIG>,
COMPILE_FEATURES,
COMPILE_FLAGS,
COMPILE_FLAGS,
COMPILE_OPTIONS,
COMPILE_OPTIONS,
COMPILE_PDB_NAME,
COMPILE_PDB_NAME_<CONFIG>,
COMPILE_PDB_OUTPUT_DIRECTORY,
COMPILE_PDB_OUTPUT_DIRECTORY_<CONFIG>,
COST,
CPACK_DESKTOP_SHORTCUTS,
CPACK_NEVER_OVERWRITE,
CPACK_PERMANENT,
CPACK_STARTUP_SHORTCUTS,
CPACK_START_MENU_SHORTCUTS,
CPACK_WIX_ACL,
CROSSCOMPILING_EMULATOR,
CXX_EXTENSIONS,
CXX_STANDARD,
CXX_STANDARD_REQUIRED,
C_EXTENSIONS,
C_STANDARD,
C_STANDARD_REQUIRED,
DEBUG_CONFIGURATIONS,
DEBUG_POSTFIX,
DEFINE_SYMBOL,
DEFINITIONS,
DEPENDS,
DEPLOYMENT_REMOTE_DIRECTORY,
DISABLED_FEATURES,
ECLIPSE_EXTRA_NATURES,
ENABLED_FEATURES,
ENABLED_LANGUAGES,
ENABLE_EXPORTS,
ENVIRONMENT,
EXCLUDE_FROM_ALL,
EXCLUDE_FROM_ALL,
EXCLUDE_FROM_DEFAULT_BUILD,
EXCLUDE_FROM_DEFAULT_BUILD_<CONFIG>,
EXPORT_NAME,
EXTERNAL_OBJECT,
EchoString,
FAIL_REGULAR_EXPRESSION,
FIND_LIBRARY_USE_LIB64_PATHS,
FIND_LIBRARY_USE_OPENBSD_VERSIONING,
FOLDER,
FRAMEWORK,
FRAMEWORK_VERSION,
Fortran_FORMAT,
Fortran_FORMAT,
Fortran_MODULE_DIRECTORY,
GENERATED,
GENERATOR_FILE_NAME,
GLOBAL_DEPENDS_DEBUG_MODE,
GLOBAL_DEPENDS_NO_CYCLES,
GNUtoMS,
HAS_CXX,
HEADER_FILE_ONLY,
HELPSTRING,
IMPLICIT_DEPENDS_INCLUDE_TRANSFORM,
IMPLICIT_DEPENDS_INCLUDE_TRANSFORM,
IMPORTED,
IMPORTED_CONFIGURATIONS,
IMPORTED_IMPLIB,
IMPORTED_IMPLIB_<CONFIG>,
IMPORTED_LINK_DEPENDENT_LIBRARIES,
IMPORTED_LINK_DEPENDENT_LIBRARIES_<CONFIG>,
IMPORTED_LINK_INTERFACE_LANGUAGES,
IMPORTED_LINK_INTERFACE_LANGUAGES_<CONFIG>,
IMPORTED_LINK_INTERFACE_LIBRARIES,
IMPORTED_LINK_INTERFACE_LIBRARIES_<CONFIG>,
IMPORTED_LINK_INTERFACE_MULTIPLICITY,
IMPORTED_LINK_INTERFACE_MULTIPLICITY_<CONFIG>,
IMPORTED_LOCATION,
IMPORTED_LOCATION_<CONFIG>,
IMPORTED_NO_SONAME,
IMPORTED_NO_SONAME_<CONFIG>,
IMPORTED_SONAME,
IMPORTED_SONAME_<CONFIG>,
IMPORT_PREFIX,
IMPORT_SUFFIX,
INCLUDE_DIRECTORIES,
INCLUDE_DIRECTORIES,
INCLUDE_REGULAR_EXPRESSION,
INSTALL_NAME_DIR,
INSTALL_RPATH,
INSTALL_RPATH_USE_LINK_PATH,
INTERFACE_AUTOUIC_OPTIONS,
INTERFACE_COMPILE_DEFINITIONS,
INTERFACE_COMPILE_FEATURES,
INTERFACE_COMPILE_OPTIONS,
INTERFACE_INCLUDE_DIRECTORIES,
INTERFACE_LINK_LIBRARIES,
INTERFACE_POSITION_INDEPENDENT_CODE,
INTERFACE_SOURCES,
INTERFACE_SYSTEM_INCLUDE_DIRECTORIES,
INTERPROCEDURAL_OPTIMIZATION,
INTERPROCEDURAL_OPTIMIZATION,
INTERPROCEDURAL_OPTIMIZATION_<CONFIG>,
INTERPROCEDURAL_OPTIMIZATION_<CONFIG>,
IN_TRY_COMPILE,
IOS_INSTALL_COMBINED,
JOB_POOLS,
JOB_POOL_COMPILE,
JOB_POOL_LINK,
KEEP_EXTENSION,
LABELS,
LABELS,
LABELS,
LANGUAGE,
LIBRARY_OUTPUT_DIRECTORY,
LIBRARY_OUTPUT_DIRECTORY_<CONFIG>,
LIBRARY_OUTPUT_NAME,
LIBRARY_OUTPUT_NAME_<CONFIG>,
LINKER_LANGUAGE,
LINK_DEPENDS,
LINK_DEPENDS_NO_SHARED,
LINK_DIRECTORIES,
LINK_FLAGS,
LINK_FLAGS_<CONFIG>,
LINK_INTERFACE_LIBRARIES,
LINK_INTERFACE_LIBRARIES_<CONFIG>,
LINK_INTERFACE_MULTIPLICITY,
LINK_INTERFACE_MULTIPLICITY_<CONFIG>,
LINK_LIBRARIES,
LINK_SEARCH_END_STATIC,
LINK_SEARCH_START_STATIC,
LISTFILE_STACK,
LOCATION,
LOCATION,
LOCATION_<CONFIG>,
MACOSX_BUNDLE,
MACOSX_BUNDLE_INFO_PLIST,
MACOSX_FRAMEWORK_INFO_PLIST,
MACOSX_PACKAGE_LOCATION,
MACOSX_RPATH,
MACROS,
MAP_IMPORTED_CONFIG_<CONFIG>,
MEASUREMENT,
MODIFIED,
NAME,
NO_SONAME,
NO_SYSTEM_FROM_IMPORTED,
OBJECT_DEPENDS,
OBJECT_OUTPUTS,
OSX_ARCHITECTURES,
OSX_ARCHITECTURES_<CONFIG>,
OUTPUT_NAME,
OUTPUT_NAME_<CONFIG>,
PACKAGES_FOUND,
PACKAGES_NOT_FOUND,
PARENT_DIRECTORY,
PASS_REGULAR_EXPRESSION,
PDB_NAME,
PDB_NAME_<CONFIG>,
PDB_OUTPUT_DIRECTORY,
PDB_OUTPUT_DIRECTORY_<CONFIG>,
POSITION_INDEPENDENT_CODE,
POST_INSTALL_SCRIPT,
PREDEFINED_TARGETS_FOLDER,
PREFIX,
PRE_INSTALL_SCRIPT,
PRIVATE_HEADER,
PROCESSORS,
PROJECT_LABEL,
PUBLIC_HEADER,
REPORT_UNDEFINED_PROPERTIES,
REQUIRED_FILES,
RESOURCE,
RESOURCE_LOCK,
RULE_LAUNCH_COMPILE,
RULE_LAUNCH_COMPILE,
RULE_LAUNCH_COMPILE,
RULE_LAUNCH_CUSTOM,
RULE_LAUNCH_CUSTOM,
RULE_LAUNCH_CUSTOM,
RULE_LAUNCH_LINK,
RULE_LAUNCH_LINK,
RULE_LAUNCH_LINK,
RULE_MESSAGES,
RUNTIME_OUTPUT_DIRECTORY,
RUNTIME_OUTPUT_DIRECTORY_<CONFIG>,
RUNTIME_OUTPUT_NAME,
RUNTIME_OUTPUT_NAME_<CONFIG>,
RUN_SERIAL,
SKIP_BUILD_RPATH,
SKIP_RETURN_CODE,
SOURCES,
SOURCE_DIR,
SOVERSION,
STATIC_LIBRARY_FLAGS,
STATIC_LIBRARY_FLAGS_<CONFIG>,
STRINGS,
SUFFIX,
SYMBOLIC,
TARGET_ARCHIVES_MAY_BE_SHARED_LIBS,
TARGET_MESSAGES,
TARGET_SUPPORTS_SHARED_LIBS,
TEST_INCLUDE_FILE,
TIMEOUT,
TIMEOUT_AFTER_MATCH,
TYPE,
TYPE,
USE_FOLDERS,
VALUE,
VARIABLES,
VERSION,
VISIBILITY_INLINES_HIDDEN,
VS_CONFIGURATION_TYPE,
VS_DEPLOYMENT_CONTENT,
VS_DEPLOYMENT_LOCATION,
VS_DESKTOP_EXTENSIONS_VERSION,
VS_DOTNET_REFERENCES,
VS_DOTNET_TARGET_FRAMEWORK_VERSION,
VS_GLOBAL_<variable>,
VS_GLOBAL_KEYWORD,
VS_GLOBAL_PROJECT_TYPES,
VS_GLOBAL_ROOTNAMESPACE,
VS_GLOBAL_SECTION_POST_<section>,
VS_GLOBAL_SECTION_PRE_<section>,
VS_IOT_EXTENSIONS_VERSION,
VS_IOT_STARTUP_TASK,
VS_KEYWORD,
VS_MOBILE_EXTENSIONS_VERSION,
VS_SCC_AUXPATH,
VS_SCC_LOCALPATH,
VS_SCC_PROJECTNAME,
VS_SCC_PROVIDER,
VS_SHADER_ENTRYPOINT,
VS_SHADER_FLAGS,
VS_SHADER_MODEL,
VS_SHADER_TYPE,
VS_STARTUP_PROJECT,
VS_WINDOWS_TARGET_PLATFORM_MIN_VERSION,
VS_WINRT_COMPONENT,
VS_WINRT_EXTENSIONS,
VS_WINRT_REFERENCES,
VS_XAML_TYPE,
WILL_FAIL,
WIN32_EXECUTABLE,
WINDOWS_EXPORT_ALL_SYMBOLS,
WORKING_DIRECTORY,
WRAP_EXCLUDE,
XCODE_ATTRIBUTE_<an-attribute>,
XCODE_EXPLICIT_FILE_TYPE,
XCODE_LAST_KNOWN_FILE_TYPE,
XCTEST
        },
	 sensitive=false,
	 morecomment=[l]{\#},
	 morestring=[b]",
	 morestring=[d]',
	}[keywords,comments,strings]

\usepackage{hyperref}
\usepackage{cleveref}

\begin{document}
%
\title{Preparing Ginkgo for AMD GPUs --\\ A Testimonial on Porting CUDA 
Code to HIP}
\titlerunning{Preparing Ginkgo for AMD GPUs}

\author{Yuhsiang M. Tsai\inst{1}\orcidID{0000-0001-5229-3739} 
\and
Terry Cojean\inst{1}\orcidID{0000-0002-1560-921X} 
\and
Tobias Ribizel\inst{1}\orcidID{0000-0003-3023-1849} 
\and
Hartwig Anzt\inst{1,2}\orcidID{0000-0003-2177-952X}}
\authorrunning{Y. Tsai et al.}
%
\institute{
Karlsruhe Institute of Technology, Karlsruhe, Germany
\and
University of Tennessee, Innovative Computing Lab, Knoxville, TN, USA
\email{firstname.lastname@kit.edu}}

\maketitle

\begin{abstract}
  With AMD reinforcing their ambition in the scientific high performance computing
ecosystem, we extend the hardware scope of the \gko{} linear algebra package to
feature a HIP backend for AMD GPUs. In this paper, we report and discuss the
porting effort from CUDA, the extension of the HIP framework to add missing 
features such as cooperative groups, the performance price of compiling HIP 
code for AMD architectures, and the design of a library providing native 
backends for NVIDIA and AMD GPUs while minimizing code duplication by using a 
shared code base.

\end{abstract}

\section{Introduction}
Over the last decade years, GPUs have been established as the main powerhouse 
in leadership
supercomputers. GPUs have proven valuable components to accelerate computations 
not only for machine learning workloads, but also for numerical 
linear algebra libraries powering computational science. As of today, AMD and 
NVIDIA are considered the main GPU manufacturers. 
In the past, software efforts primarily focused on NVIDIA GPUs due to the 
comprehensive CUDA development environment and the common adoption in HPC 
centers. With the next leadership supercomputers deployed in the US National 
Laboratories being equipped with AMD GPUs, and the US Exascale Computing 
Project's mission to provide math library functionality on the leadership 
systems, we extend the scope of the \gko library to feature an AMD GPU backend.

In this paper, we report and discuss the effort of porting a CUDA-focused 
library to the HIP ecosystem. We elaborate on the use of the perl-based script
provided by AMD that aims at simplifying the transition process, its pitfalls
and flaws. We also assess the performance HIP-based code achieves on NVIDIA
architectures when compiled using NVIDIA's \texttt{nvcc} compiler.

Transitioning a code base from one architecture to another, and platform 
portability in general, is an important problem in the software technology 
ecosystem. In particular, the number of adopters and contributors of community 
software scales only in the presence of good platform portability. 
The effort of porting a software stack to new architectures is, for example,  
described for molecular dynamics 
algorithm in~\cite{Kuznetsov2019-rc}, and for the solution of finite 
element problems in~\cite{Zubair2019-nr}. Concerning performance portability, 
the authors of~\cite{Sun2018-bx} compare the algorithm performance for CUDA, 
HC++, HIP, and OpenCL backends.

Compared to previous work, we highlight that this work contains the following 
novel contributions:
\begin{itemize}
    \item We discuss the porting of linear algebra 
    kernels from CUDA to HIP.
    \item We add technology to the HIP ecosystem that is lacking but needed, 
    e.g., a subwarp cooperative group concept with shuffle operations.
    \item We compare the performance of HIP and CUDA kernels coming from the 
    same code base and providing the same functionality.
    \item Up to our knowledge, \gko is the first open-source sparse linear 
    algebra library supporting several matrix types (Coo, Csr, Sellp, Ell, 
    Hybrid), solvers (CG, BiCG, GMRES, etc), preconditioner (block-jacobi) and 
    factorization (ParILU and ParILUT) on AMD and NVIDIA GPUs.
    \item We ensure full result reproducibility by archiving all performance 
    results.
\end{itemize}

Before providing more details about the porting effort in \Cref{sec:porting}, 
we recall 
some background information about CUDA and HIP in \Cref{sec:related}. We 
present the 
results of the experiments of the same kernels being compiled by CUDA and HIP in
\Cref{sec:experiments}. We conclude in \Cref{sec:conclusion} with a 
summary of this paper.

\section{Background}
\label{sec:related}
\subsection{Compute Unified Device Architecture - CUDA}
NVIDIA developed the CUDA programming model and the corresponding \texttt{nvcc} 
compiler enabling developers to write GPU-parallel programs using the C or C++ 
programming language. Also, NVIDIA provides several math libraries, like 
cuBLAS, cuSPARSE, and cuSOLVER containing ready-to-use numerical algorithms and 
core functionalities allowing users to easily develop a parallel application 
without writing device kernel functions.

In \Cref{code:cuda_example}, 
CUDA uses \kernel as the declaration specifier to
tell the compiler this function runs on a GPU and uses execution configuration
syntax (\lcudasyntex\rcudasyntex) to represent the configuration of grid and 
block dimensions, execution stream, and dynamically-sized shared memory.
Moreover, developers can provide additional information at compile-time to 
optimize the execution performance like \launchbound to limit 
the register usage. 


\begin{lstlisting}[language=C,caption={CUDA kernel launch 
syntax.},label={code:cuda_example},morekeywords={constexpr, hipLaunchKernelGGL,
HIP_KERNEL_NAME}]
template <int value>
__global__ void dummy_kernel(const int num, int *__restrict__ array) {
  // kernel_code
}
int main() {
  // allocation of memory and calculation of grid/block_size
  dummy_kernel<4> <<<dim3(grid_size), dim3(block_size)>>>(num, array);
  return 0;
}
\end{lstlisting}

\subsection{C++ Heterogeneous-Compute Interface for Portability - HIP}
As a counterpart to NVIDIA's CUDA ecosystem, AMD more recently developed the GPU
compute programming language and library ecosystem ``RadeonOpenCompute'' (ROCm).
ROCm is the first open-source HPC platform for GPU computing shipping with
several math libraries, like rocBLAS, rocSPARSE, rocSOLVER, etc. This enables
users to develop GPU-ready applications in ROCm like in the CUDA ecosystem.

Aside from ROCm, AMD also provides a HIP abstraction that can be seen as a 
higher layer on top of the ROCm ecosystem, enveloping also the CUDA ecosystem. 
The idea behind HIP is to increase platform portability of software by 
providing an interface through which functionality of both, ROCm and
CUDA can be accessed. Obviously, this would remove the burden of converting or 
rewriting code for different hardware architectures, therewith also reducing 
the maintenance effort for libraries supporting several backends.

In \Cref{code:hip_example}, HIP uses the same declaration specifier \kernel
like CUDA, but a different execution configuration syntax.
HIP handles kernels featuring template parameters with the macro 
HIP\_KERNELS\_NAME. 
Although HIP 
also provides the \launchbound flag for kernel optimization, the effect differs 
from the CUDA ecosystem due to the architectural differences between AMD and 
NVIDIA GPUs. 
\begin{lstlisting}[language=C,caption={HIP kernel launch 
syntax.},label={code:hip_example},morekeywords={constexpr, hipLaunchKernelGGL,
HIP_KERNEL_NAME}]
template <int value>
__global__ void dummy_kernel(const int num, int *__restrict__ array) {
  // kernel_code
}
int main() {
  // allocation of memory and calculation of grid/block_size
  hipLaunchKernelGGL(HIP_KERNEL_NAME(dummy_kernel<4>), dim3(grid_size),
                     dim3(block_size), 0, 0, num, array);
  return 0;
}
\end{lstlisting}

\subsection{Difference between AMD and NVIDIA GPUs}
The primary technical difference between AMD and NVIDIA GPUs is the number of 
threads 
that are executed simultaneously in a wavefront/warp. In NVIDIA GPUs, a
warp contains 32 threads, in AMD GPUs, a wavefront contains 64 threads. This 
difference potentially impacts all other parameter configurations and has to be 
taken into account when designing kernels and setting thread block size, 
shared memory and register usage, and compute grid size for valid parameter 
settings and optimal kernel performance. 

Less relevant for the kernel design and parameter choice is that AMD and NVIDIA 
GPUs differ in the number of multiprocessors accumulated in a single device and
in the memory bandwidth. While these are still relevant for kernel 
optimization, they
rarely impact the correctness of a kernel design. We elaborate on the 
optimization 
of kernel parameters in \Cref{sec:porting_workflow}.

As of today, AMD's ROCm ecosystem -- and the HIP development ecosystem -- still 
lacks some key functionality of the CUDA ecosystem. For example, HIP lacks a 
cooperative group interface that can be used for flexible thread programming 
inside 
a wavefront, see \Cref{sec:cooperative_groups}. 

\section{Porting CUDA functionality to the HIP ecosystem}
\label{sec:porting}
Next, we report and discuss how we ported \gko{}'s GPU functionality available 
for CUDA backends to the HIP ecosystem. To understand the technical 
realization, it is however useful to first elaborate on \gko{}'s design.

\subsection{\gko design}
\begin{figure}[!t]
  \centering
  \includegraphics[width=0.9\linewidth]{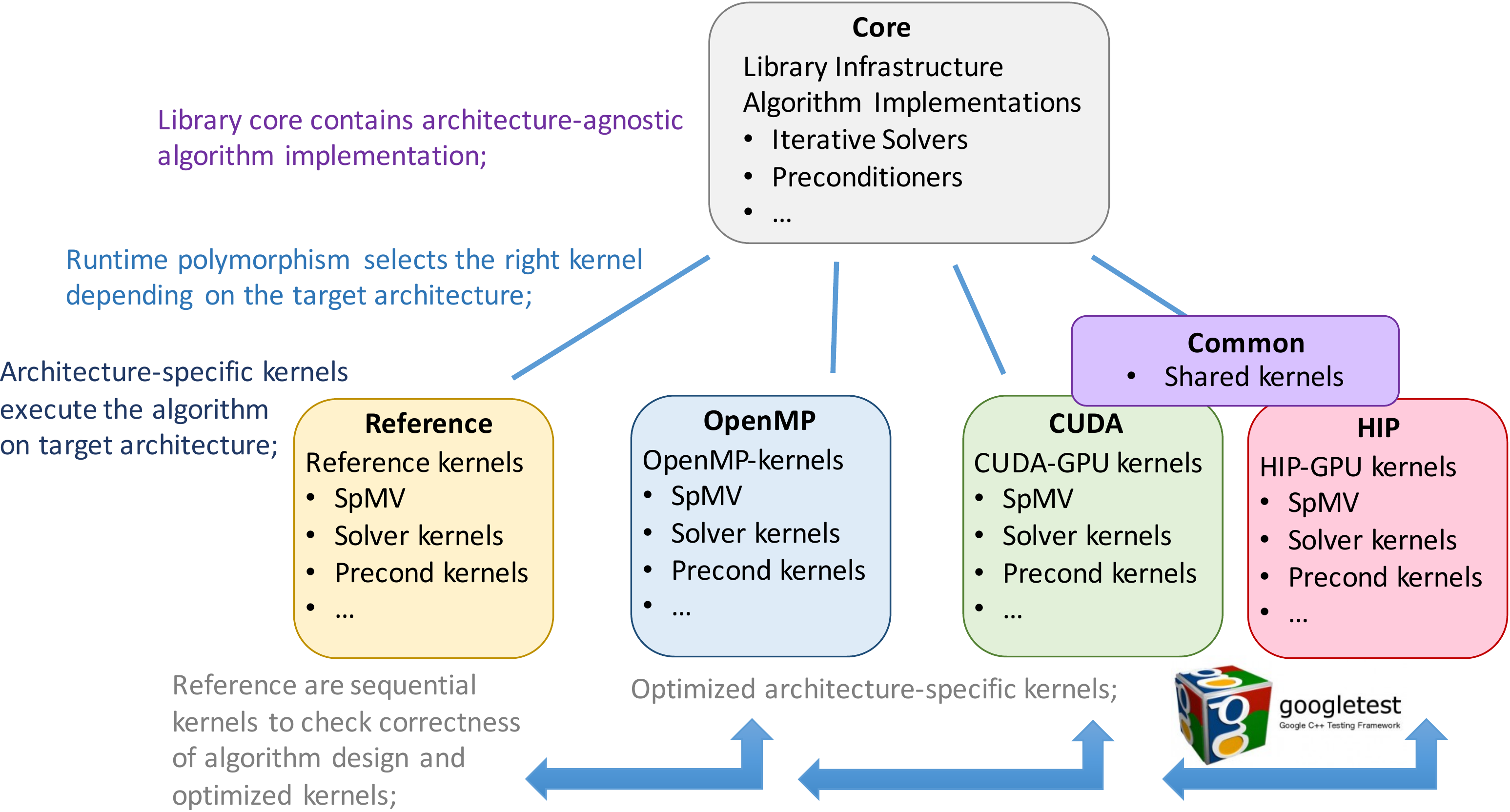}
  \caption{The \gko library design overview. The components added when
    extending the scope to AMD GPUs are the ``HIP'' and the ``Common'' modules.}
  \label{fig:gko_desgin}
\end{figure}

A high-level overview of \gko{}'s software architecture is visualized in 
\Cref{fig:gko_desgin}. The library design collects all classes and generic 
algorithm skeletons in the ``core'' library which, however, is useless without 
the driver kernels available in the ``omp'', ``cuda'', and ``reference'' 
folders. We note that ``reference'' contains sequential CPU kernels used to 
validate the correctness of the algorithms and as reference implementation for 
the unit tests realized using the googletest\cite{googletest} framework.
The ``include'' folder contains the public interface. Extending 
\gko{}'s scope to AMD architectures, we add the ``hip'' folder containing the 
kernels in the HIP language, and the ``common'' folder for platform-portable 
kernels with the intention to reduce code duplication, see 
\Cref{sec:avoid_duplication}.

To reduce the effort of porting \gko to AMD architectures, we use the same 
base components of \gko{} like \texttt{config}, \texttt{binding},  
\texttt{executor}, \texttt{types} and \texttt{operations},
which we only extend and adapt to support HIP.
\begin{itemize}
\item \texttt{config}: hardware-specific information like warp size, 
lane\_mask\_type, etc.;
\item \texttt{binding}: the C++ style overloaded interface to vendors' BLAS and sparse BLAS
library and the exception calls of the kernels not implemented;
\item \texttt{executor}: the ``handle'' controlling the kernel execution and the ability 
to switch the execution space (hardware backend);
\item \texttt{types}: the type of kernel variables and the conversion between library 
variables and kernel variables;
\item \texttt{operations}: a class aggregating all the possible kernel implementations
  such as reference, omp, cuda and hip, which allows to switch between
  implementations at runtime.
\end{itemize}

Moreover, some components are not officially supported by vendors, e.g. complex 
number atomic\_add\footnote{A complex atomic\_add involves separate real 
and imaginary atomic\_add and thus is not strictly an atomic operation, as 
no ordering between the individual components of multiple complex atomic 
operations is guaranteed.} on CUDA and HIP, and warp-wide cooperative groups on 
HIP. For the functionality missing in both vendor 
ecosystems, we implement CUDA device functions providing the functionality and 
apply the work flow listed in \Cref{alg:gko_porting_workflow} to generate 
corresponding HIP kernels.
For components missing only in one vendor ecosystem, we implement kernels
providing the same functionality in the other ecosystem.
In particular, as the HIP ecosystem currently lacks the warp-wide cooperative 
groups we make heavy use of, we implement device functions that provides this
functionality for AMD architectures, see \Cref{sec:cooperative_groups}.

\subsection{Avoiding code duplication}
\label{sec:avoid_duplication}

Despite the fact that the HIP ecosystem allows to compile the kernels for both 
AMD and NVIDIA GPUs, we currently plan to still provide native support in the 
CUDA ecosystem. This choice is motivated by the wider adoption of CUDA in the 
high performance computing community on the one side, and the unclear future of 
this functionality remaining in the HIP ecosystem on the other side. A third 
reason is that preserving native CUDA support allows to utilize novel 
CUDA-specific technology, e.g., dynamic parallelism.
Extending \gko to AMD GPUs, a primary goal was to avoid a significant level of 
code duplication. For this purpose, we created the ``common'' folder containing 
all kernels and device functions that are identical or the CUDA and the HIP 
executor except for kernel configuration parameters (such as warp size or 
\texttt{launch\_bounds}). These configuration parameters are not set in the 
kernel file contained in the ``common'' folder, but in the files located in 
``cuda'' and ``hip'' that are interfacing these kernels. This way we can avoid 
code duplication while still configuring the parameters for optimal kernel 
performance on the distinct hardware backends.

\subsection{Cooperative groups}
\label{sec:cooperative_groups}
CUDA 9 introduced cooperative groups for flexible thread programming.
Cooperative groups provide an interface to handle thread block and warp groups
and apply the shuffle operations that are used heavily in \gko for optimizing
sparse linear algebra kernels. HIP~\cite{hip} only supports block and grid
groups with \texttt{thread\_rank()}, \texttt{size()} and \texttt{sync()}, but no
subwarp-wide group operations like shuffles and vote operations.

For enabling full platform portability, a small codebase, and preserving the 
performance of the optimized CUDA kernels, we implement cooperative group 
functionality for the HIP ecosystem. Our implementation supports the 
calculation of size/rank and shuffle/vote operations 
inside subwarp groups. We acknowledge that our cooperative group 
implementation may not support all features of CUDA's cooperative group 
concept, but all functionality we use in \gko.

The cross-platform cooperative group functionality we implement with shuffle 
and vote operations covers CUDA's native implementation. HIP only interfaces 
CUDA's warp operation without \texttt{\_sync} suffix (which refers to 
deprecated functions), so we use CUDA's native warp operations to avoid 
compiler warning and complications on NVIDIA GPUs with compute capability 7.x 
or higher. 
We always use subwarps with contiguous threads, so we can use the block index to
identify the threads' subwarp id and its index inside the subwarp. 
We define
\begin{scriptsize}
\[
  \begin{split}
    \texttt{Size} &= \text{Given subwarp size}\\
    \texttt{Rank} &= \texttt{tid \% Size}\\
    \texttt{LaneOffset} &= \lfloor{} \texttt{tid \% warpsize / Size} 
\rfloor{}\times \texttt{Size}\\
    \texttt{Mask} &= \texttt{$\sim 0$ >> (warpsize - Size) <<
    LaneOffset}
  \end{split}
\]
\end{scriptsize}

where \texttt{tid} is local thread id in a thread block such that \texttt{Rank} 
gives the local id of this subwarp, and \texttt{$\sim 0$} is a bitmask of 32/64 
bits, same bits as \texttt{lane\_mask\_type}, filled with 1 bits according to 
CUDA/AMD architectures, respectively. Using this definition, we can realize the 
cooperative group interface, for example for the \texttt{shfl\_xor}, \texttt{ballot}, 
\texttt{any}, and \texttt{all} functionality:
\begin{scriptsize}
\[
  \begin{split}
    \texttt{subwarp.shfl\_xor(data, bitmask)} &= \texttt{\_\_shfl\_xor(data, 
    bitmask, Size)}\\
    \texttt{subwarp.ballot(predicate)} &= \texttt{(\_\_ballot(predicate) \& 
    Mask) >> LaneOffset} \\
    \texttt{subwarp.any(predicate)} &= \texttt{(\_\_ballot(predicate) \& Mask) 
    != 0 }\\
    \texttt{subwarp.all(predicate)} &= \texttt{(\_\_ballot(predicate) \& Mask) 
    == Mask} 
  \end{split}
\]
\end{scriptsize}
Note that we use the \texttt{ballot} operation to implement \texttt{any} and 
\texttt{all} operations. The original warp \texttt{ballot} returns the answer for the 
entire warp, so we need to shift and mask the bits to access the subwarp 
results. The \texttt{ballot} operation is often used in conjunction with bit 
operations like the population count (\emph{popcount}), which are provided by 
C-style type-annotated intrinsics \texttt{\_\_popc[ll]} in CUDA and HIP. To 
avoid any issues with the 64bit-wide lane masks on AMD GPUs, we provide a 
single function \texttt{popcnt} with overloads for 32 and 64 bit integers as 
well as an architecture-agnostic \texttt{lane\_mask\_type} that provides the 
correct (unsigned) integer type to represent a (sub)warp lane mask.


\begin{lstlisting}[language=diff,caption={reduce kernel. Green part is cooperative 
group implementation, and red part is legacy implementation},label={code:reduce}]
 template <int Size, typename ValueType>
 __global__ void reduce(ValueType *__restrict__ data, int inner_loops) {
   auto local_data = data[threadIdx.x];
   for (int i = 0; i < inner_loops; i++) {
+    auto group = group::tiled_partition<Size>(group::this_thread_block());
     #pragma unroll
-    for (int bitmask = 1; bitmask < Size; bitmask <<= 1) {
+    for (int bitmask = 1; bitmask < group.size(); bitmask <<= 1) {
-      const auto remote_data = __shfl_xor(local_data, bitmask, Size);
+      const auto remote_data = group.shfl_xor(local_data, bitmask);
       local_data = local_data + remote_data;
     }
   }
   data[threadIdx.x] = local_data;
 }
\end{lstlisting}

\begin{figure}[!h]
  \centering
  \begin{tabular}{lcr}
    \includegraphics[width=0.56\linewidth]{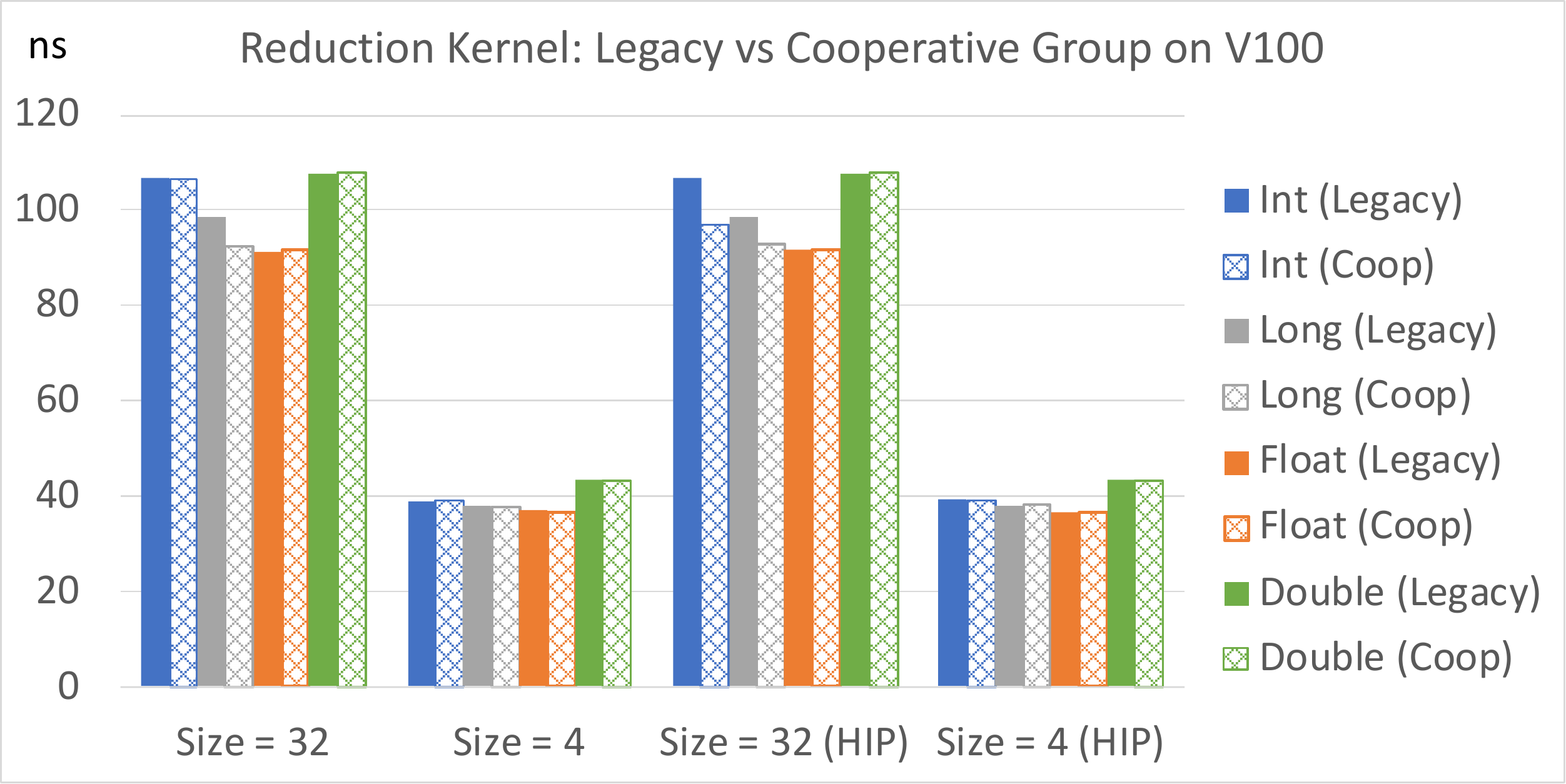}
    & \quad\ \quad &
    \includegraphics[width=0.37\linewidth]{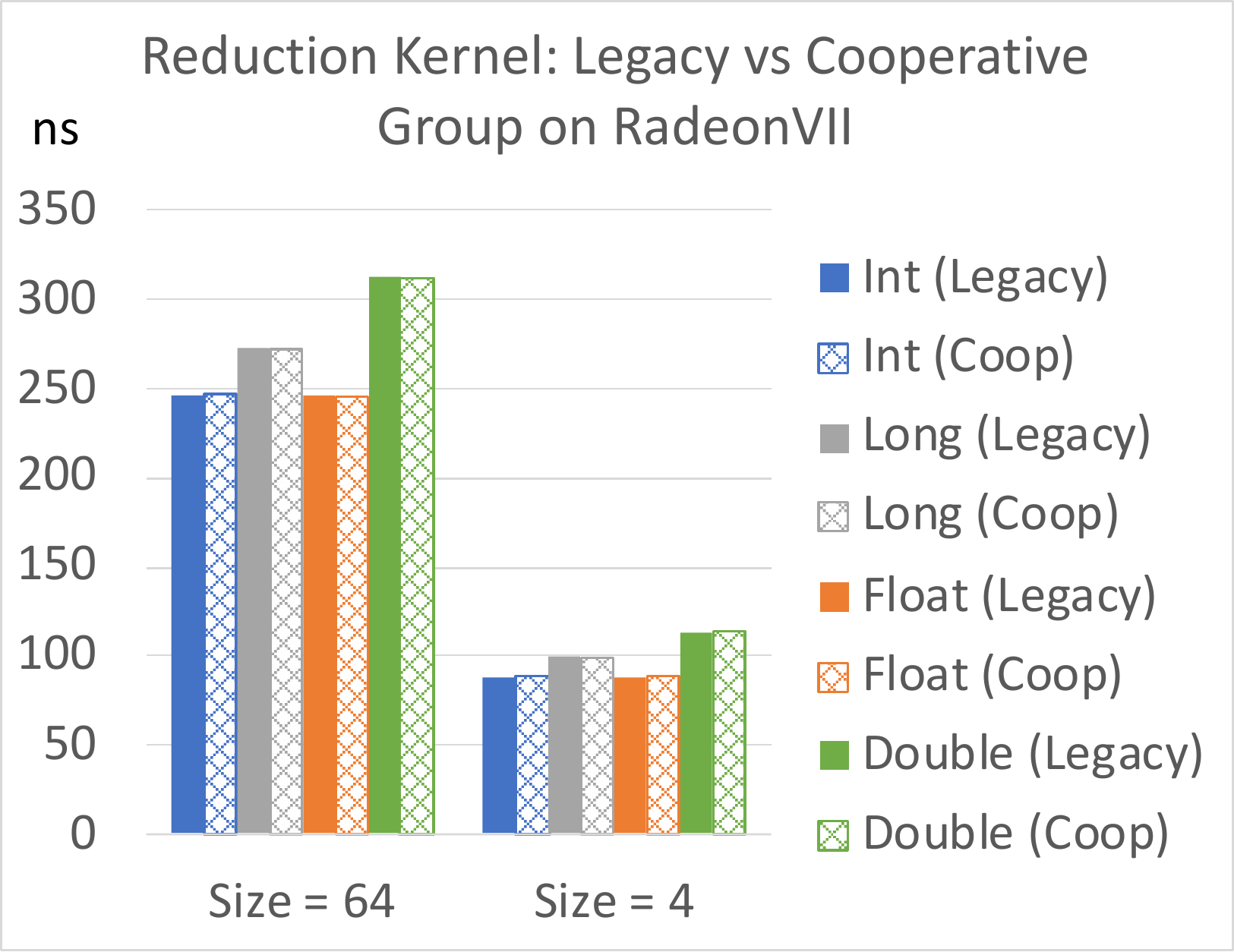}
  \end{tabular}
  \caption{\gko's cooperative groups vs. legacy functions for different data 
  types on V100 (left) and RadeonVII (right).}
  \label{fig:legacy_vs_coop}
\end{figure}

To assess the performance of our cross-platform cooperative group
implementation, we use the local reduction kernel shown in 
\Cref{code:reduce} that utilized either the vendor's legacy functionality (red) 
or \gko's cross-platform cooperative group interface (green). 
In \Cref{fig:legacy_vs_coop}, we report the runtime needed for 100 reduction 
operations (after a warm-up phase of 10 reductions) on NVIDIA's V100 GPU and 
AMD's RadeonVII GPU. 
To exclude the overhead of the kernel launch and memory operations, we run the 
kernel executing ``inner\_loops'' reductions (line~4 of \Cref{code:reduce}) for 
``inner\_loops = 1000'' and ``inner\_loops = 2000'' and report the runtime 
difference. 
This way, we can isolate the runtime needed for the warp-wide reduction by 
excluding the overhead of the kernel launch and memory operations.
The results
identifies \gko's cross-platform cooperative group implementation as
competitive to the vendor's native implementation.
Both implementations use the same strategy for the reduction operation, and 
both implementations execute the reduction loop (line~7-12 of 
\Cref{code:reduce}) exactly $\log_{2}(Size)$ times.
For the execution time for different values of $Size$, the theoretical 
performance ratios are $\frac{\log_2(4)}{\log_2(64)} = 0.333$ on the RadeonVII 
and $\frac{\log_2(4)}{\log_2(32)} = 0.4$ on the V100. In the experimental 
evaluation, we observe average ratios $\frac{\mathrm{runtime}(Size = 
4)}{\mathrm{runtime}(Size = 64)} = 0.360$ and $\frac{\mathrm{runtime}(Size = 
4)}{\mathrm{runtime}(Size = 32)} = 0.394$ for the RadeonVII and the V100 GPUs, 
respectively.

\subsection{Porting via the Cuda2Hip script}
\label{sec:cuda2hip}

For easy conversion of CUDA code to the HIP language, we use a script based on
the hipify-perl script provided by AMD with several modifications to meet our
specific needs. First, the script generates the target filename including the 
path in the ``hip'' directory. Then AMD's hipify-perl script is invoked to 
translate the CUDA kernels to the HIP language, including the
transformation of NVIDIA's proprietary library functions to AMD's library
functions and the kernels launch syntax. Next, the script changes all 
CUDA-related header, namespace, type and function names to the corresponding 
HIP-related names. By default, the script hipify-perl fails 
to handle namespace definitions. For example, the hipify-perl script changes
\texttt{namespace::kernel\lcudasyntex...\rcudasyntex(...)} to
\texttt{namespace::hipLaunchKernelGGL(kernel, ...)}, while the correct
output would be \texttt{hipLaunchKernelGGL(namespace::kernel, ...)}.
Thus, the script ultimately needs to correct the namespaces generated by the 
hipify-perl script.

\subsection{Porting workflow}
\label{sec:porting_workflow}
In \Cref{alg:gko_porting_workflow}, we sketch the workflow we use for porting
\gko{}'s CUDA backend to HIP.
Step~1 introduces a set of variables to represent the architecture-specific  
parameters such as the warp size (32 on CUDA devices, 64 on AMD devices) and 
optimization parameters. 
Step~2 moves the identical kernel codes into the ``common'' folder we 
introduced in \Cref{sec:avoid_duplication}. 
We include the code in the ``common'' folder after setting the configuration 
variables in Step~3 and Step~4. Step~5 runs the script Cuda2Hip script detailed 
in \Cref{sec:cuda2hip} to generate the corresponding hip files. Ultimately, we 
modify the hip ``config'' file in Step~6. 
After completion of these
steps, the validity and correctness of the porting effort is tested. This is
realized by invoking \gko{}'s unit test framework that employs googletest to
check the correctness of the high performance kernels -- in particular also the
CUDA and HIP backends -- against the reference kernels.

We note that \gko's cross-platform cooperative group extension presented in
\Cref{sec:cooperative_groups} dramatically reduces backend-specific 
implementations and allows to use a shared kernel in ``common'' for both, the 
NVIDIA and the HIP backend.

\begin{algorithm}[!h]
    \caption{\gko{}'s porting workflow}
    \label{alg:gko_porting_workflow}
    \begin{algorithmic}[1]
        \STATE Use a variable to represent the architecture-specific parameters
        \STATE Move all shared code into a ``common'' file
        \STATE Set the architecture-specific parameters before including a 
        ``common'' file
        \STATE Include the ``common'' file
        \STATE Use the Cuda2Hip script for converting the code
        \STATE Modify the hip file ``config'' to support different architectures
    \end{algorithmic}
\end{algorithm}



\subsection{Porting statistics for \gko}
\label{sec:statistics}
\begin{figure}
\begin{tabularx}{\textwidth}{lX}
    \includegraphics[width=.65\columnwidth]{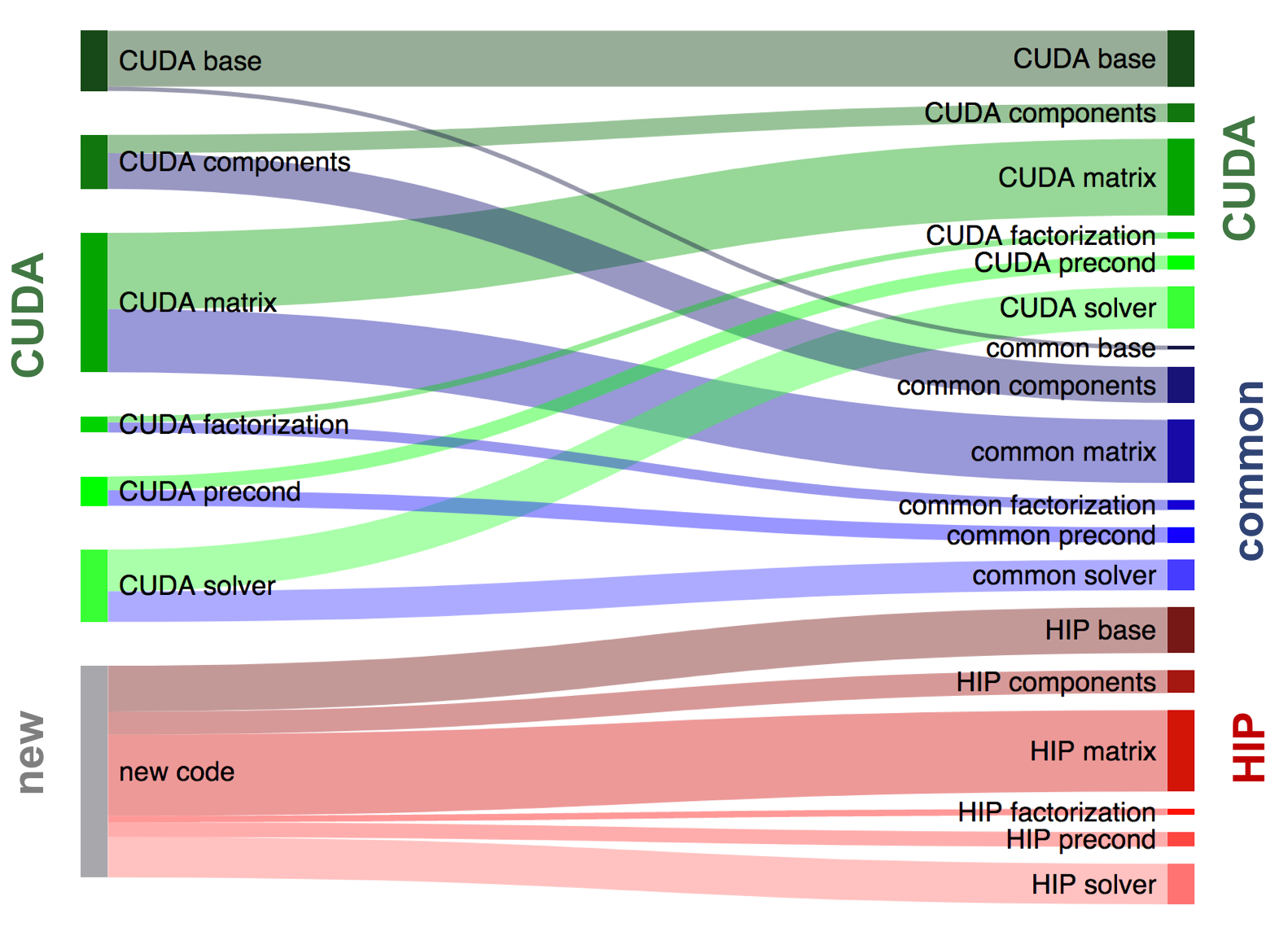} &
\vspace{-4cm}
\begin{scriptsize}
    \begin{tabular}{lrrr}
      \hline
      \hline
      Module &  common & cuda & hip\\
      \hline
      base            & 112	  &   1435	& 1176   \\
      component       & 919	  &   467	    &  589   \\
      matrix          & 1617    &   1908	& 2048   \\
      factor	      & 262	  &   159	    &  165   \\
      precond		  & 395	  &   356	    &  375   \\
      solver          & 780	  &   1071	& 1038   \\
      \hline
      \hline
    \end{tabular}
\end{scriptsize}
\end{tabularx}
    \label{fig:tryboth}
    \caption{Left: Reorganization of the \gko library to provide a HIP
      backend for AMD GPUs. Right: (Physical) Lines of code in the ``common'', 
      ``cuda'', 
      and ``hip'' modules of the \gko library, ignoring the unit tests.}
\end{figure}

With the setup and tools described, extending the scope of \gko to cover also 
AMD GPUs is a smooth process. We acknowledge that some kernels that are 
heavily tuned for performance needed additional attention, most notably the 
multiprecision block-Jacobi kernel~\cite{adptiveprecisionblockjacobi}. Aside 
from this, the addition of the HIP ecosystem required slight modifications to 
the library architecture, most importantly the addition of the ``common'' 
module containing the kernels that are identical up to parameter settings for 
the CUDA and the HIP ecosystems.
In the left figure of \Cref{fig:tryboth}, we visualize how existing code lines are relocated and 
new code lines are added when extending \gko{}'s scope to support also HIP.
The exact number of code lines contained in the distinct modules of the 
extended \gko library are listed in the right table of \Cref{fig:tryboth}. We note that about one 
third of the code base is shared between the CUDA and the HIP executor, and 
that by creating the ``common'' folder we actually avoided duplicating 4,000 
lines of code. 
The other modules each contain about 5,000 lines of code. While most submodules 
are comparable in size, the more significant differences for ``base'' and 
``component'' stem from the differing comprehensiveness of the ecosystems and 
possibilities of architecture-specific optimization.


\vspace{-.5cm}

\section{Experiments}
\label{sec:experiments}

To assess how well the HIP ecosystem interfaces to the CUDA technology, we 
compare HIP code compiled for NVIDIA GPUs with native CUDA code. More 
precisely, we apply the 
porting workflow we described in \Cref{sec:porting} to high performance sparse 
linear algebra kernels of \gko{}'s CUDA backend, and compare the performance of 
the generated HIP code when being compiled for NVIDIA GPUs with the original 
kernel performance. 
We run our experiments on NVIDIA's V100 (SXM2 16 GB) \cite{volta} with cuda 
9.2.148 and hip 3.1.20044-3684ef8 (which is the latest version on Jan. 31 
2020). We compare the Sellp, Coo, and cuSPARSE/hipSPARSE (Splib\_Csr) SpMV 
kernels, and the 
Conjugate Gradient Solver employing the Sellp SpMV kernel for the Krylov 
subspace generation using either CUDA and HIP on the same device. 
For result reproducibility, we archive all performance results in a public
repository 
\footnote{\url{https://github.com/ginkgo-project/ginkgo-data/tree/V100_cuda_hip}}.
We evaluate the performance of the \gko SpMV for more than 2,800 matrices 
from the SuiteSparse Matrix Collection \cite{suitesparse}. 
We run two iterations for warm-up and ten iterations to obtain average 
performance values. 

\begin{figure}
  \vspace{-0.5cm}
    \centering
\begin{tabular}{lcr}
    \includegraphics[width=0.45\linewidth]{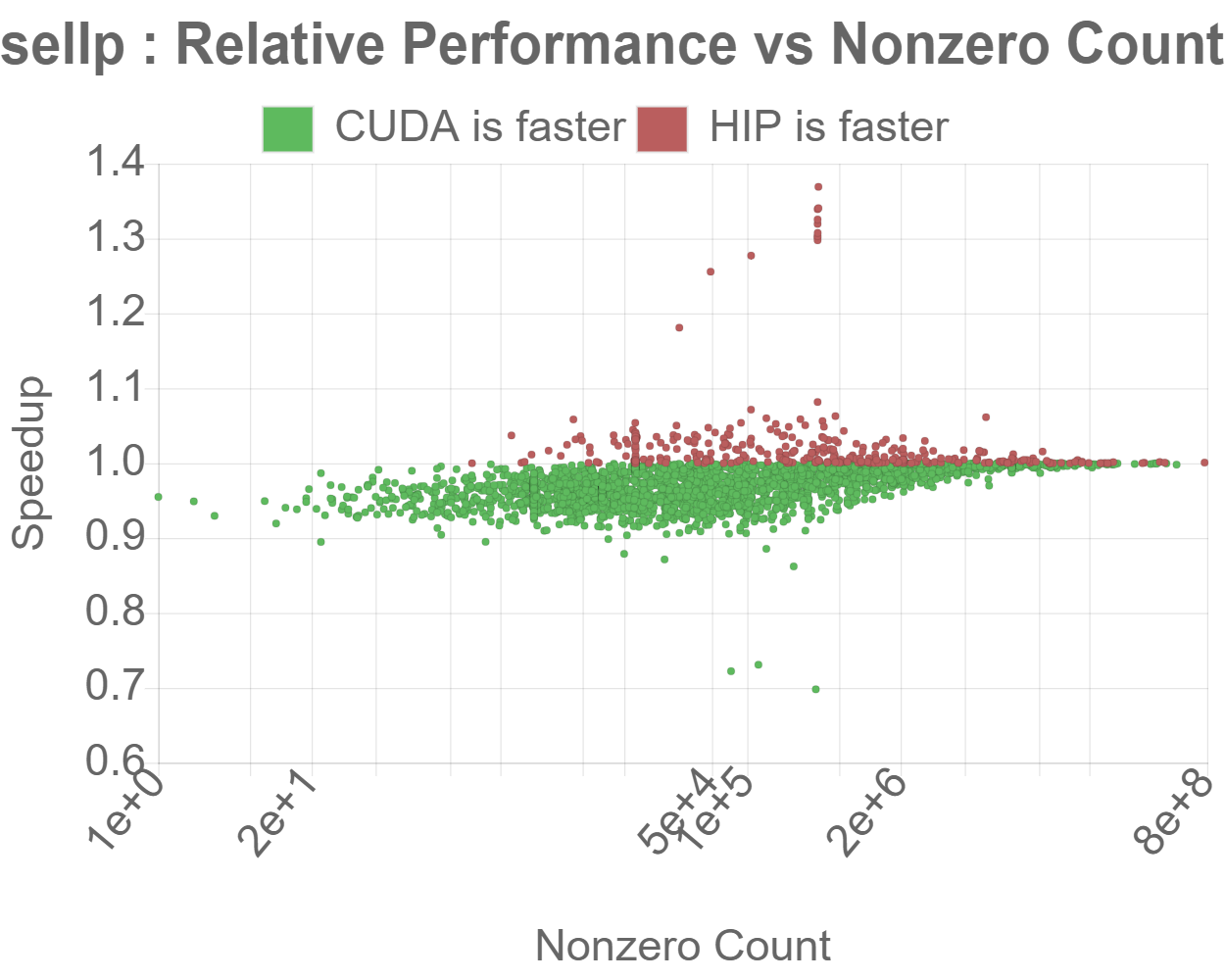}
& \quad\ \quad &
    \includegraphics[width=0.45\linewidth]{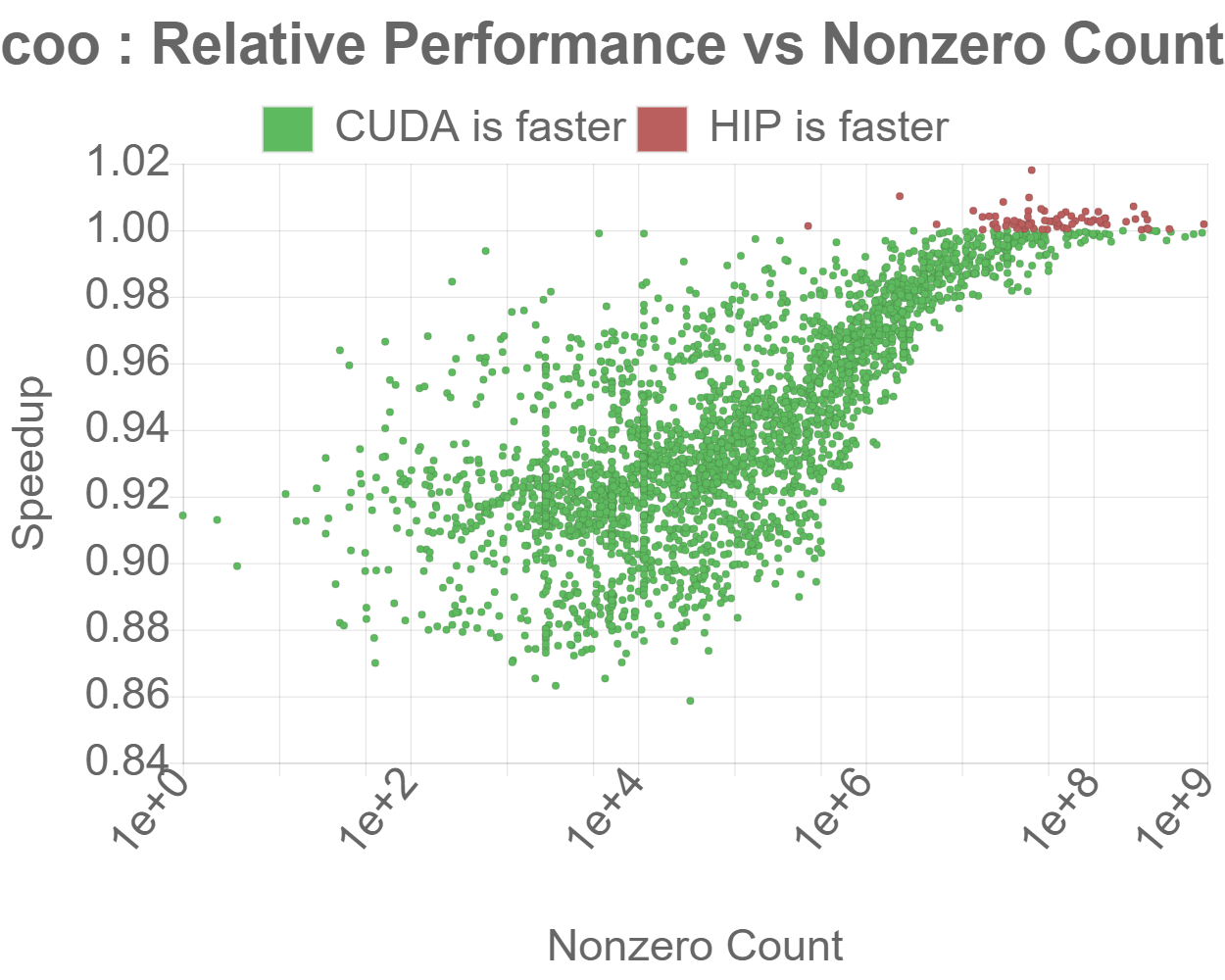}
\end{tabular}
    \caption{Sellp SpMV (left) and Coo SpMV implemented in CUDA or HIP.}
    \label{fig:sellp_relative_performance}
  \vspace{-0.5cm}
\end{figure}

On the left-hand side of \Cref{fig:sellp_relative_performance}, we evaluate the 
performance for \gko{}'s Sellp SpMV kernel, which does not use atomic 
operations. On the right-hand side of \Cref{fig:sellp_relative_performance}, we 
do the same comparison for \gko{}'s Coo SpMV kernel which does rely on atomic 
operations.
Running on NVIDIA's V100 GPU, one would expect to see small overhead of the 
HIP code interfacing CUDA code compared to native CUDA code. While this may 
prove 
true for most problems, we see some outliers where using the native CUDA 
implementation results in significant performance benefits. Surprisingly, for 
some test cases the HIP kernels achieve significantly better performance -- 
even though HIP ultimately compiles with NVIDIA's \texttt{nvcc} compiler.
The generated PTX code indicates that the differences may be attributed 
to slightly different types of \texttt{load} instructions being emitted, which 
in turn use different caches.

\begin{figure}
  \vspace{-0.5cm}
    \centering
\begin{tabular}{lcr}
    \includegraphics[width=0.45\linewidth]{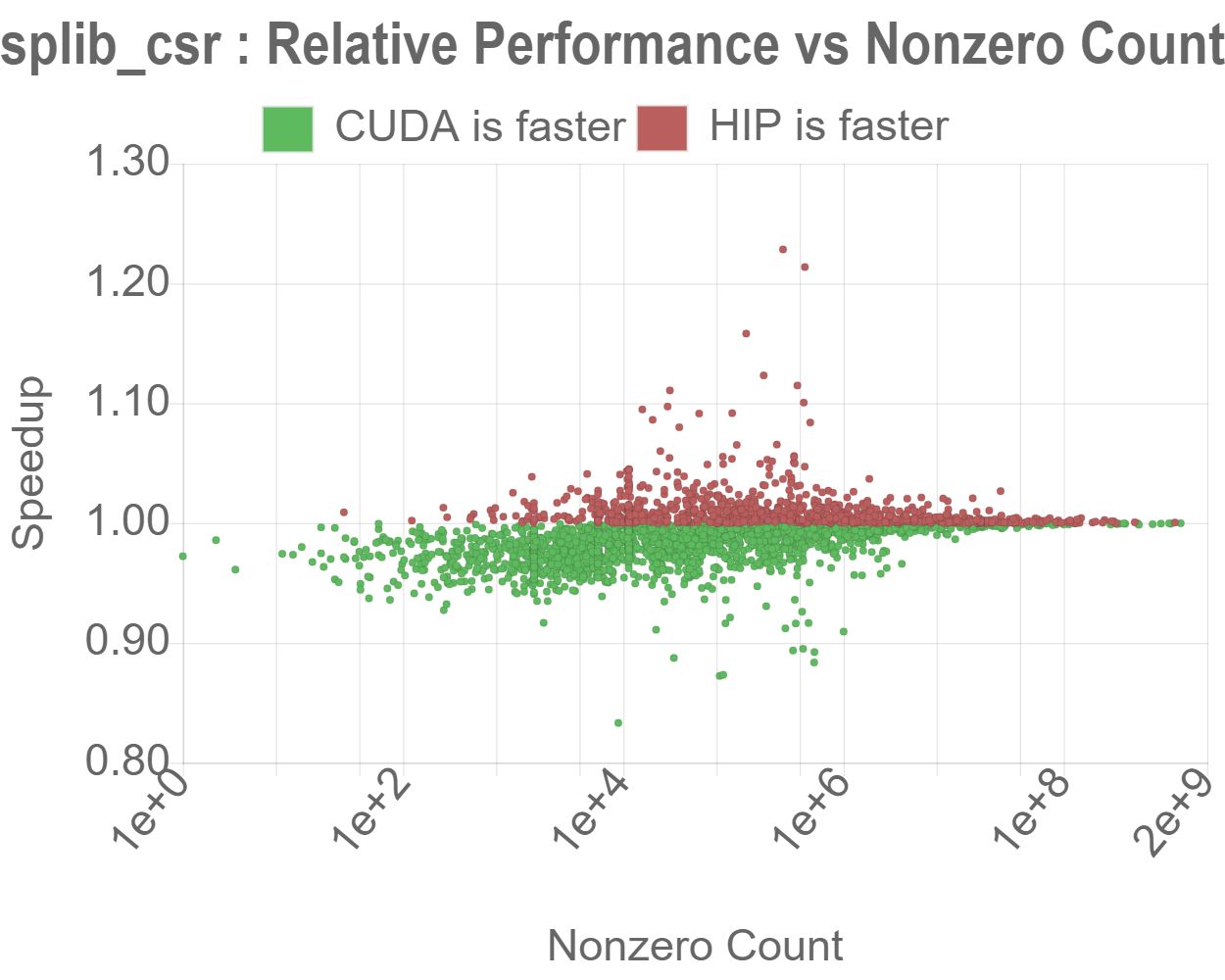}
& \quad\ \quad &
    \includegraphics[width=0.45\linewidth]{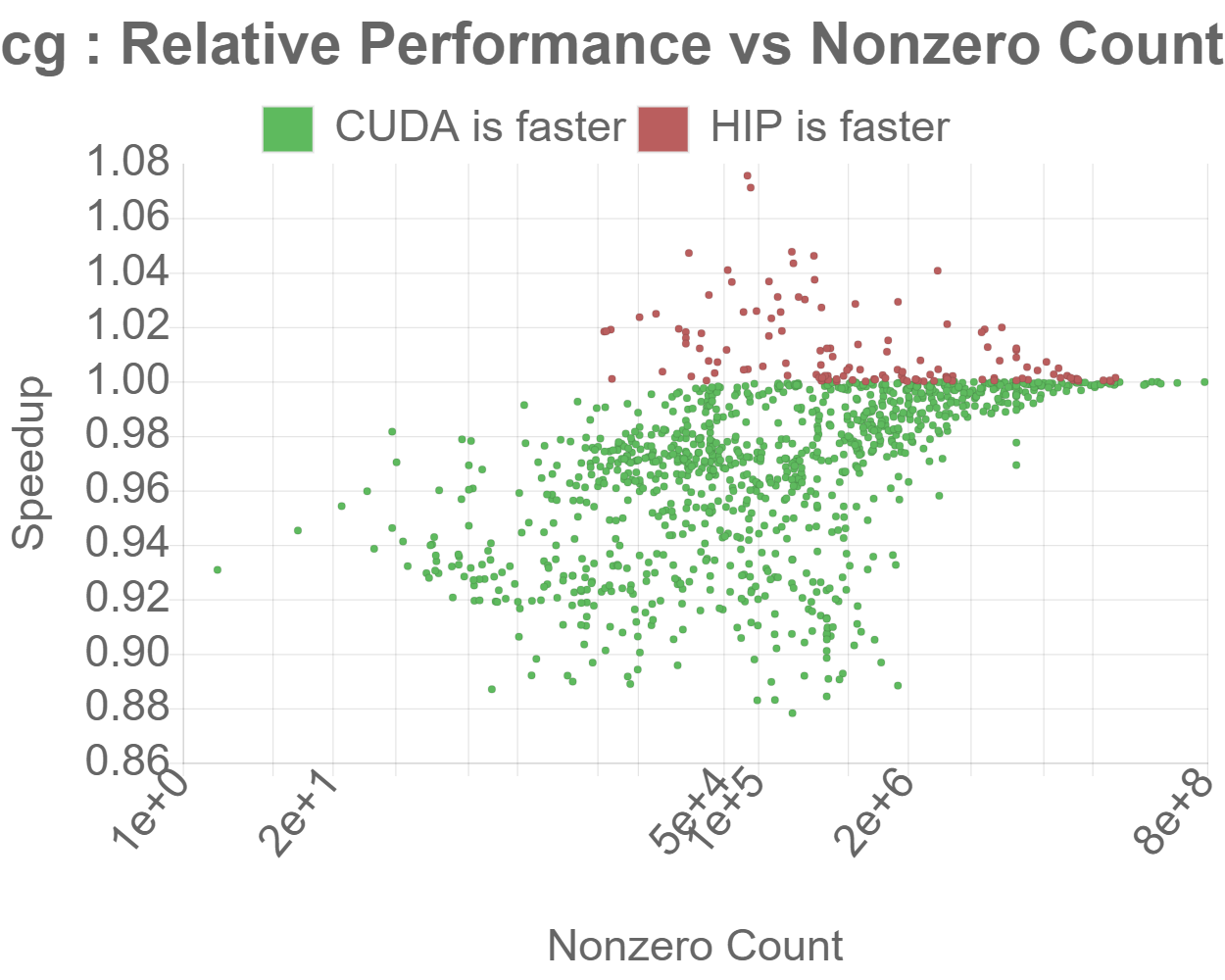}
\end{tabular}
    \caption{Performance comparison for vendors' Csr SpMV (left) and 1,000 
    iterations of \gko{}'s CG solver (right).}
    \label{fig:cg_relative_performance}
  \vspace{-0.5cm}
\end{figure}

In \Cref{fig:cg_relative_performance} we do the same experiment for the 
vendors' Csr SpMV (left-hand side) and 1,000 iterations of \gko{}'s Conjugate
Gradient (CG) solver using \gko{}'s Sellp SpMV (right-hand side).
For the vendors' Csr SpMV comparison on the left, the performance 
differences reflect only the overhead of the invocation of cuSPARSE by 
hipSPARSE. 
In the CG performance comparison on the right, we observe up to 15\% 
performance degradation coming from the aforementioned differences in code 
generation. This is in accordance with Philip C. Roth~\cite{Roth2019-eh} 
who compares the performance of CUDA and HIP for the scalable heterogeneous 
computing (SHOC) benchmark~\cite{Danalis2010-re}. 

\begin{figure}[!h]
  \vspace{-0.3cm}
    \centering
\begin{tabular}{lcr}
    \includegraphics[width=0.45\linewidth]{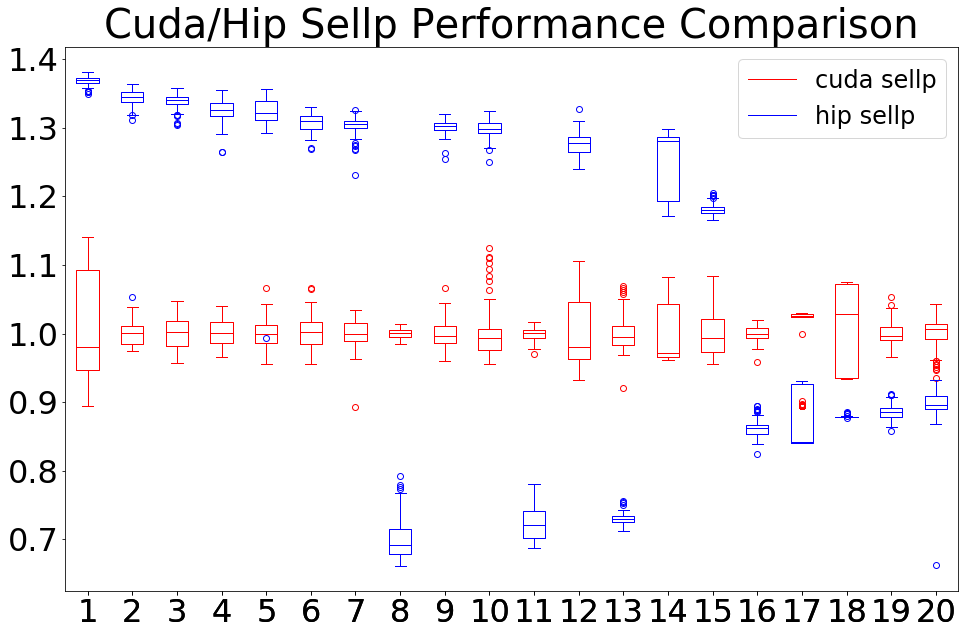}
& \quad\ \quad &
    \includegraphics[width=0.45\linewidth]{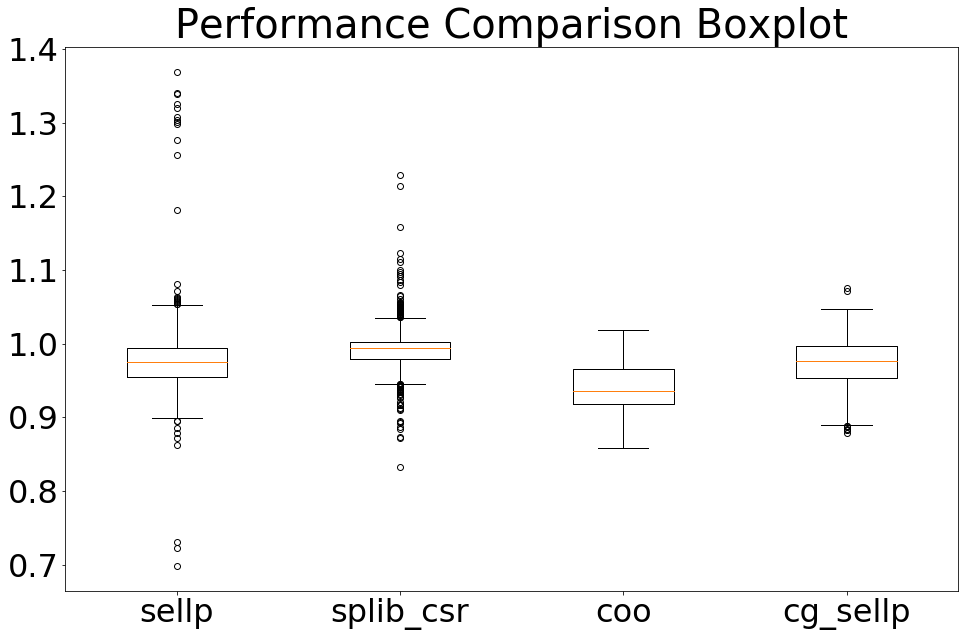}
\end{tabular}
    \caption{Left: Performance variance for outliers in Sellp
    SpMV kernel analysis (\Cref{fig:sellp_relative_performance}). All
    performance is normalized to the mean CUDA performance, CUDA performance in
    red, (relative) AMD performance in blue. Right: Performance statistics for
    all test cases and all kernels/algorithms.}
    \label{fig:statistics}
  \vspace{-0.3cm}
\end{figure}

As some of the performance differences in \Cref{fig:sellp_relative_performance} 
are significant, we investigate in \Cref{fig:statistics} (left) the mean and 
variance of the 20 most significant outliers in the Sellp SpMV analysis in 
\Cref{fig:sellp_relative_performance} (left). These statistics are collected 
from over 20 runs, each averaging the kernel characteristics over 100 
invocations.
Acknowledging the reproducibility of these outliers, we emphasize that they are 
still almost negligible when considering the complete test suite of more than 
2,800 test matrices: The performance ratio statistics on the right-hand side of 
\Cref{fig:statistics} reveal that the performance means for all functionalities 
are just slightly below 1.0. Furthermore, 50\% of the test cases show less than 
3\% performance difference, and 90\% of the test cases show less than 10\% 
performance difference. This reveals that HIP introduces only negligible 
overhead when comparing to CUDA-native code.





\section{Conclusion}
\label{sec:conclusion}
We elaborated how we extend the hardware scope of the \gko
linear algebra package to feature a HIP backend for AMD GPUs. We
discussed the porting effort, and how the use of a shared code base reduces to 
minimize code duplication in a library providing native backends for NVIDIA and 
AMD GPUs. 
We also detailed the addition of functionality currently lacking in the HIP 
ecosystem and evaluated the performance price of compiling HIP code for NVIDIA 
architectures. We found that a significant portion of sparse linear algebra 
kernels allows for good platform portability.
In future, we will create a Intel GPU backend and compare 
the porting process with the HIP backend integration.
\section*{Acknowledgment}
This research was supported by the Exascale Computing Project (17-SC-20-SC) and 
the Helmholtz Impuls und Vernetzungsfond VH-NG-1241.


\FloatBarrier
\bibliographystyle{splncs04}
\bibliography{references}

\end{document}